\documentclass[acmsmall]{acmart}

\AtBeginDocument{%
  \providecommand\BibTeX{{%
    \normalfont B\kern-0.5em{\scshape i\kern-0.25em b}\kern-0.8em\TeX}}}

\setcopyright{acmcopyright}
\copyrightyear{2018}
\acmYear{2018}
\acmDOI{10.1145/1122445.1122456}

\begin{document}

\title{XR-Ed Framework: Designing Instruction-driven and Learner-centered Extended Reality Systems for Education}

\author{Kexin Yang}
\email{kexiny@andrew.cmu.edu}
\affiliation{%
  \institution{Carnegie Mellon University}
  \streetaddress{5000 Forbes Ave}
  \city{Pittsburgh}
  \state{PA}
  \country{USA}
  \postcode{15213}
}

\author{Xiaofei Zhou}
\affiliation{%
  \institution{University of Rochester}
  \streetaddress{250 Hutchison Rd}
  \city{Rochester}
  \state{NY}
  \country{USA}
 }
\email{xzhou50@ur.rochester.edu}

\author{Iulian Radu}
\affiliation{%
  \institution{Harvard University}
  \streetaddress{13 Appian Way}
  \city{Cambridge}
  \state{MA}
  \country{USA}}
\email{iulian_radu@g.harvard.edu}


\begin{abstract}
Recently, the HCI community has seen an increased interest in applying Virtual Reality (VR), Augmented Reality (AR) and Mixed Reality (MR) into educational settings. Despite many literature reviews, there still lacks a clear framework that reveals the different design dimensions in educational Extended Reality (XR) systems. Addressing this gap, we synthesize a broad range of educational XR to propose the XR-Ed framework, which reveals design space in six dimensions (Physical Accessibility, Scenario, Social Interactivity, Agency, Virtuality Degree, Assessment). Within each dimension, we contextualize the framework using existing design cases. Based on the XR-Ed Design framework, we incorporated instructional design approaches to propose XR-Ins, an instruction-oriented, step-by-step guideline in educational XR instruction design. Jointly, they aim to support practitioners by revealing implicit design choices, offering design inspirations as well as guide them to design instructional activities for XR technologies in a more instruction-oriented and learner-centered way.
\end{abstract}



\maketitle

\section{Introduction}
Virtual technologies can improve students’ academic performance and motivation, e.g. \cite{barsom2016systematic,chen2017review,hamilton2020immersive,kaminska2019virtual,kommetter2019pedagogical,maas2020virtual, martin2017virtual,merchant2014effectiveness, parmaxi2020augmented,radianti2020systematic,radu2014augmented,yilmaz2016educational}, students’ social and collaborative skills \cite{kaufmann2005general,martin2010design}, and students’ psychomotor and cognitive skills \cite{feng2009Trends}. Among them Virtual Reality (VR) and Augmented Reality (AR) have long been a popular design space for educational technology, and recently, Mixed Reality (MR) also increasingly applied for educational use \cite{radianti2020systematic}. These immersive technologies have the potential to increase learner motivation and engagement \cite{martin2017virtual}, promote a full student-centered learning experience \cite{winn2002research}, support collaborative and situated learning and enable learners to more concretely and tangibly access previously physically inaccessible/invisible content \cite{wu2013current}.

In recent years, multiple literature reviews that explore AR, VR or MR applications in education emerged, e.g., \cite{barsom2016systematic, chen2017review, hamilton2020immersive, kaminska2019virtual, kommetter2019pedagogical, maas2020virtual, martin2017virtual, merchant2014effectiveness, parmaxi2020augmented, radianti2020systematic, radu2014augmented, yilmaz2016educational}, mostly focused on one of the three technologies with specific research questions. These immersive technologies have similarity as well as  nuanced differences, recent efforts starting to provide a more comprehensive review across the use of AR, VR, MR educational applications \cite{kommetter2019pedagogical, maas2020virtual}, which encourage us to review on all the different types of extended reality (XR) \cite{mann2018all}. Existing reviews and studies have pointed out that many educational affordances of XR can be achieved by other multimedia, and in order to maximize its unique educational benefits, the core of their success lies in integration with effective instructional design.

Though many systematic reviews have been published, there is still a lack of a clear framework that both reveals different design dimensions existing in XR systems and organizes them in a learner-centered way. Previous work argued instructional design is more important than other compelling features of XR technologies, and should be placed at the center of the design process of educational XR application \cite{merchant2014effectiveness, parmaxi2020augmented, wu2013current}.

In the field of education there has been lots of work done on understanding how to teach properly (with and without technology). While XR is getting used more in education, many XR practitioners (e.g. technology designers and developers) may not be intimately familiar with educational theory and instructional design; so most reviews could have limited use in practice. To reduce the chances of XR design and development efforts being "hit-and-miss, driven by intuition and ‘common-sense’ extrapolations”, and be more solidly underpinned by research-informed models and frameworks \cite{dalgarno2010learning}, conceptual support from the perspectives of instructional design and learning theory are needed, for educational technology practitioner (i.e. researchers, designers) \cite{ertmer2012teacher, hamilton2020immersive, maas2020virtual, radianti2020systematic}, especially given the fact that many XR practitioners may not be intimately familiar with education.

Therefore, our literature review aims at gaining a new lens in educational XR to answer the research questions: 1) When designing XR for education, what design dimensions should be considered by practitioners? 2) What instruction-centered design guidelines and procedures could practitioners follow when they design educational XR applications? To answer these two research questions, we reviewed 70 papers in the domain of XR systems for education use, mainly from recent 5 years.

Our paper contributes the XR-Ed framework, which maps out six design dimensions in the format of spectrums instantiated with specific XR systems, that XR practitioners could consider when designing and implementing their own apps. We hope, through making the design space more transparent, it would help practitioners situate their XR systems on the spectrums when making design choices and weighing design trade-offs. Based on the framework, we further proposed XR-Ins design guidelines, a set of instruction-centered XR education technology design guidelines to facilitate designers and researchers to both make good use of XR design affordances and emphasize effective instruction/learning experience to design in a more structured way. This guideline was informed by backward design, a known theoretical approach in instructional design, involving “Identify desired results”, “Determine acceptable evidence” and “Plan learning experience and instruction” \cite{wiggins2005understanding}.

\section{Related Work}

\subsection{Distinguishing and defining AR, VR and MR}
Virtual Reality (VR) refers to a whole simulated reality built with computer systems using digital formats to create a realistic immersive experience \cite{martin2017virtual}, and can be defined as “hardware and software systems that seek to perfect an all inclusive, sensory illusion of being present in another environment”  \cite{radianti2020systematic}. 
Augmented Reality (AR) has been defined more diversely \cite{wu2013current}. Usually, AR technologies superpose synthetic elements like 3D objects, multimedia contents or text information onto real-world images \cite{hsieh2011conceptual}. From the standpoint of experience it affords users, AR can provide users technology-mediated immersive experiences in which real and virtual worlds are blended, in which “a real world context is dynamically overlaid with coherent location or context sensitive virtual information”. From a technology standpoint, \cite{milgram1995augmented} define AR as a form of virtual reality where the participant’s head-mounted display is transparent, allowing a clear view of the real world” \cite{milgram1995augmented} From the standpoint of features it supports, Azuma defines AR as a system that fulfills three basic features: a combination of real and virtual worlds, real-time interaction and accurate 3D registration of virtual and real objects. 
Mixed Reality (MR) came into existence later than AR and VR, and can be defined as a situation where real world and virtual world objects are presented together. As situated in the Reality-Virtuality continuum proposed by \cite{milgram1995augmented}), mixed reality can consist of two main ideas: augmented reality (AR) and augmented virtuality (AV), which will be explained further later in Dimension 5 in the XR-Ed framework.

\subsection{Existing Reviews and Framework on XR in Education
}
Previous reviews generally revealed the trends, advantages, limitations and the vision of educational VR or AR applications. \cite{akccayir2017advantages, chen2017review, freina2015literature, garzon2019systematic, kaminska2019virtual, wu2013current,yilmaz2016educational}. For example, \cite{hamilton2020immersive, radu2014augmented,merchant2014effectiveness} compared learning outcomes and benefits in VR versus non-VR \cite{hamilton2020immersive}, AR versus non-AR \cite{radu2014augmented}, and different VR-based instruction \cite{merchant2014effectiveness}. \cite{radianti2020systematic} took a similar instruction-driven approach in its literature review, with a different target (VR in higher-ed settings). Some reviews happens in a particular knowledge domain, for example, \cite{barsom2016systematic,arici2019research,parmaxi2020augmented} investigated how AR technologies support medical professional training \cite{barsom2016systematic}, science education \cite{arici2019research} and language learning \cite{parmaxi2020augmented}.

Existing literature proposed several frameworks for AR, VR or MR in education. \cite{radianti2020systematic} proposed a framework of VR for education, including 14 fine-detailed design elements that practitioners should consider in the design process, such as realistic surroundings, passive observations, etc. \cite{wu2013current} proposed a framework that identified three viable instructional approaches for AR: in the design of AR instructional activities or systems, practitioners could choose to emphasize “roles”, “location” or “task”, each relate to AR affordance (e.g. emphasizing “roles” may increase learners’ sense of presence, immediacy and immersion; while emphasizing “tasks” may potentially enhance authenticity). For MR, \cite{kommetter2019pedagogical} developed a pedagogical framework to evaluate MR use in education including dimensions of 1) type (AR, VR); 2) operation site (school, home); 3) teaching method (group, single, partner); 4) level of pervasion (assisting, enhancing, replacing); 5) supervision (1:1, 1:n).

Existing frameworks still have yet to provide a complete flow, (i.e. starting from desired learning objectives to appropriate assessment results), which our XR-Ins Guideline seeks to address, by mapping a step-by-step procedure to guide the design of an effective learning experience supported by XR technologies.

\subsection{Learning Theory and Instructional Design Approach
}
In the domain of education, numerous prior work has identified evidence-based learning sciences principles to scaffold effective learning. A variety of instructional approaches were adopted the design of XR learning environments, including game-based learning \cite{rosenbaum2007location,squire2007augmented,squire2007mad},  place-based learning \cite{klopfer2008augmented,mathews2010using}, participatory simulations \cite{klopfer2008augmented, rosenbaum2007location,squire2007augmented}, problem-based learning \cite{liu2009outdoor,squire2007augmented}, role playing \cite{rosenbaum2007location}, studio-based pedagogy \cite{mathews2010using}, and jigsaw method \cite{dunleavy2009affordances}. Most prior educational XR reviews focus more on the classic learning paradigms (e.g. cognitivism, behaviorism, constructivism, connectivism, experimentalism) \cite{martin2017virtual, radianti2020systematic}. While these learning paradigms provide fundamental knowledge, they may be too theoretical to prompt immediate actions. Hereby we focus on recent, evidence-based learning sciences principles, based on the known book by Clark and Mayer on e-Learning design, to provide inspirations for XR technologies design \cite{clark2003Elearing}. Due to space limitation, we included only principles we consider most important and relevant to the XR design. 
\begin{enumerate}
    \item Multimedia Principle: Using words and graphics concurrently, rather than text exclusively can help engage both visual and auditory channels of learners. 
    \item Contiguity Principle: Words explaining a concept and its accompanying images should be presented close to each other  (spatial contiguity), and simultaneously (temporal contiguity), to facilitate learners’ retention and understanding. 
    \item Coherence Principle: Irrelevant, extraneous or inapplicable information (audio, visual, words) should be eliminated, to allow learners to concentrate on critical elements only.
    \item Modality Principle: The way we present information should be dependent on how complex the information is. (E.g. a complicated process may be more effectively conveyed in a visual format to reduce learners’ ‘information overload’).
    \item Personalization Principle: Systems that use a conversational voice or tone, rather than formal authoritative tone, can often be more relatable and engaging, and facilitate learners to process the knowledge and content.  
    \item Signaling (cueing) Principle: E-learning is more effective when cues are added that guide learners’ attention to relevant elements,or highlight the organization of the materials. 
\end{enumerate}
Numerous other books offer insight on how to design e-learning technologies \cite{horton2011learning}, or how to design effective instruction in a systematic \cite{allen2012leaving, torrance2019agile,yocum2015design} way. 
As for systematic procedure of instructional design, Understanding by Design (UbD) \cite{wiggins2005understanding}, is a known theoretical approach in this area. In this broad approach, researchers proposed Backward Design, a three-stage approach practitioners should follow for learning experience design, that consist of 1) Identify desired results; 2) Determine acceptable evidence; 3) Plan learning experiences and instruction. We seek to combine these known approaches in instructional design in the XR-Ins (XR-Instruction) Guideline, to help XR practitioners adopt an instruction-driven and learner-centered way for XR system design.

\section{Methodology}
Our search scanned in three databases, including the ACM Digital library, the dblp computer science bibliography, and Google Scholar, and used keywords: “XR”, “Virtual Reality”,“VR”, “Augmented Reality”, “AR” and “Mixed Reality”,“MR” combined with “education”, “learning”, “school”, and “classroom”. To be included in the review, the paper must be (1) from a peer-reviewed conference or journal, (2) be written in English, (3) use one of the XR technologies for educational purposes. The general review method of this paper was inspired by An and Holstein’s review on teaching augmentation systems, which propose TA framework \cite{an2020ta}.

Two researchers met weekly to discuss how XR is applied in each paper and review the analysis results of each paper. During the first round of analysis, two researchers evaluated the study and XR system design comprehensively, including through extracting the following basic information to gain initial familiarity. At a meta level, we looked at: 1) study type (and for studies being experimental, then the independent variables (IV) and dependent variables (DV) were collected), 2) educational goal, 3) learning sciences theory utilized, 4) research questions asked 5) research result and contribution, 6) number of users, 7) role of users, 8) age of users, 9) educational settings targeted (e.g. formal, informal), 10) education domain targeted (e.g. physics, math). From the perspective of design, we looked at 1) key existing system design components, 2) design recommendation. This stage of literature review served as helping researchers to get background contexts for the specific XR applications were designed for and used.

During the second round, researchers paid special attention to how XR technologies were utilized to facilitate the learning of domain knowledge, by annotating papers on this particular aspect. They distill higher-level commonality of XR usage from the specific use cases in each paper, and as a result, came up with common “\textit{mechanisms}” that signify XR’s common usage in education. The definition of “\textit{mechanism}” draws inspiration from a meta-review \cite{wu2013current}, and has connotation of the“\textit{functions and affordances} of XR in education identified in the paper”. From the initial mechanisms extracted, researchers took a bottom-up approach to conduct thematic analysis. Specifically, they cluster and group the similar “\textit{mechanism}” to more easily see the common higher-level themes that capture the essence of them. As one concrete example of this process, for instance, while \cite{allison2000virtual,keifert2017agency,lindgren2011supporting} target different domains (i.e. gorillas’ behaviors, state of matter, and asteroid movement/force), this identified a higher level mechanism that they share is that they all  learners to \textit{role-play} as an object/character related to the target knowledge, to help learners grasp the “deep structure” of the knowledge \cite{lindgren2011supporting}. Four rounds of iterative clustering and grouping, they arrived at seven mechanisms in Table.~\ref{Tab:mecha} in appendix.

Finally, by utilizing the list of collected mechanisms, two researchers worked collaboratively to conduct interpretation sessions, where they \textit{cross-check} the definition of each mechanism and if every individual XR application can be \textit{mapped back} to these extracted mechanisms to check the validity. Inspired by the presentation of An and Holstein’s TA framework, researchers found continuum (spectrums) to better capture the rich design space than the categorical mechanisms, and thus proposed a 6-dimension XR-Ed (XR-Education) framework.

\section{XR-Ed Design Framework}

In this section, we present the XR-Ed (XR for Education) framework, consisting of six dimensions that reveal a rich design space for XR application for educational use. We contextualize each dimension of the XR-Ed Framework through discussing how they are concretely instantiated in actual XR systems, and provide brief discussion and reflection for each dimension. We hope to, in doing so, reveal the key design choices that may normally remain implicit, and help XR practitioners in their process of designing and building XR applications for educational use.

\begin{figure}[htbp]
\centering
\includegraphics[width=0.85\columnwidth]{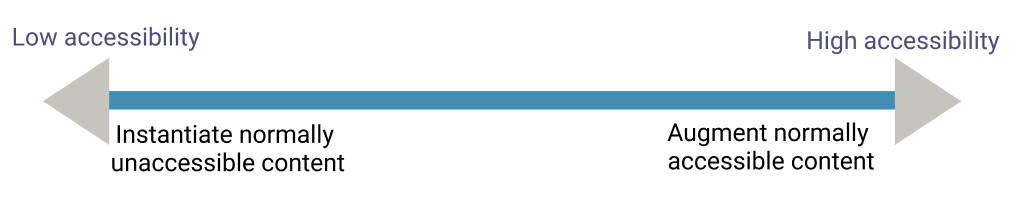}
\caption{The spectrum of D1: Physical Accessibility of Target Learning Content}
\label{fig:d1}
\end{figure}


\subsection{D1: Physical Accessibility of Learning Content -- How physically/visually accessible is the target knowledge in a normal life scenario?}

The first dimension in the XR-Ed Framework addresses how physically / visually accessible the content knowledge is in ordinary life scenarios. By “accessible”, we mean if the content to be learned (e.g. a science phenomena, medical surgery procedure) are normally observable or noticeable.

\subsubsection{Instantiate normally inaccessible content} 
One prominent affordance of XR, is to create scenarios and learning experiences that are normally inaccessible by learners, or processes that are hard to capture by human eyes, through various techniques including 3D modeling, simulation and visualization. For example, Radu and Schneider use AR to design activities where learners can visualize and interact with dynamic representations of hidden forces (e.g. visualizing electrons, magnetic fields, light or radio waves), to make challenging concepts visible and accessible to novices \cite{radu2019can}. Similarly, \cite{jiang2020augmented} visualizes abstract concepts and invisible science phenomena, to help students learn thermodynamics (9th-grade science); \cite{liu2007mixed} uses MR, enabling learners to view the virtual solar system and visualize the  process of photosynthesis, that are normally invisible / too slow for human to capture and observe.

\subsubsection{Augment normally accessible content}
Contrary to the XR applications that seek to visualize and allow learners to see or experience normally inaccessible knowledge, some applications (especially MR and AR) seek to augment the normally accessible content knowledge to be richer or more engaging. For example, \cite{hirabayashi2019development} built a MR system that teaches sign language and fingerspelling, through which learners can watch sign language and fingerspelling motion in three dimensions as if they are looking at them in the real world. This augment over the traditional media to provide learners with richer information so as they can grasp the hand gesture better through controlling the movement and rotation of the 3D model.  

\subsubsection{Discussion}
We argue for XR practitioners to consider XR’s affordances of simulating or visualizing a normally inaccessible or unobservable process, as this may offer a unique, novel learning experience to learners, which could be otherwise extremely costly or impossible to experience physically. For example, \cite{freina2015literature} argued that the main motivation for VR use is it gives the opportunity to live and experiment situations that “cannot be accessed physically” . They depict four typical scenarios that XR can afford learners to access the “normally inaccessible”: 1) Time inaccessibility: e.g. allowing learners to travel in time to experience different historical periods; 2) Physical inaccessibility: e.g. exploring the solar system by freely moving around planets; 3) Dangerous situation: e.g. providing training for firefighters in physically and psychologically stressful situation through simulation live firefighting; learning gorilla behaviors through interacting with a virtual gorilla \cite{allison2000virtual}; or using MR to simulates earthquake to teach basic physics \cite{yannier2015learning} 4) Ethic problem: for example, allowing non-experts to perform surgery. Researches found “turning the invisible visible” could trigger students’ interests and curiosities in science as well as developing students’ connection between science learning in classrooms and their lives \cite{jiang2020augmented,martin2017virtual, klopfer2008augmented}.

\begin{figure}[htbp]
\centering
\includegraphics[width=0.85\columnwidth]{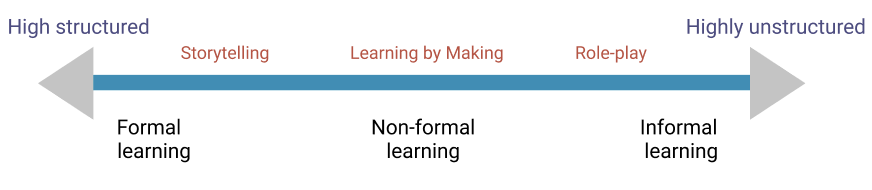}
\caption{The spectrum of D2: Formal Degree of Learning Scenario}
\label{fig:d2}
\end{figure}
\subsection{D2: How formal is the target learning scenario?}
The second dimension of the XR-Ed Framework asks how formal the learning scenarios are. Generally speaking, “formal learning” is pragmatic and organized,  and “informal learning” casual, unstructured, spontaneous and often unintentional. If informal learning comes consciously with a defined purpose, it becomes “non-formal” \cite{colardyn2004validation, coll2019formal,rogers2014base}. 

\subsubsection{Formal learning}: 

Formal learning normally refers to educational activities that are structured, goal-oriented and instructor-led \cite{colardyn2004validation,coll2019formal,rogers2014base}. XR has been used to amplify formal learning that happens in the classroom, targeting knowledge prescribed by national or provincial educational standards \cite{ayasoufi2019}. XR used in formal learning often seeks to augment normal instruction with digital media content, physical or virtual toolkit or 3D modeling. For example, in a large junior high classroom, \cite{chen2015employing} uses AR toolkit to augment instructions with real-time AR image depicting movements of sun-earth turntables, to teach earth science (i.e. phenomena of day, night and seasons). In a university classroom, first-year college students use AR-embedded books which display 3D images upon scanning, to facilitate geography learning. AR can facilitate formal learning in forms of toolkits, for example, Construct3D for mathematics and geometry learning (i.e. vector analysis, descriptive geometry, and geometry); unnamed AR toolkit to teach lever principle (physics)\cite{liu2020ar}.

\subsubsection{Non-formal Learning}
Non-formal learning normally refers to informal learning that comes with a defined purpose. It is usually the result of intentional effort, but need not follow a formal syllabus or be governed by external accreditation and assessment \cite{colardyn2004validation, coll2019formal,rogers2014base}. In this sense, educational XR applications were frequently used to provide a gamified experience, in an effort to better engage learners to learn formal learning content. Three techniques that we found that were frequently adopted include 1) Storytelling, 2) Learning by making (doing), 3) Role-play or Participatory Simulation.

\textbf{Storytelling}. Many educational XR applications operate on the idea of storytelling, to render the content knowledge more engaging, approachable or easy to understand. For example, to facilitate math learning, ARMath tells a story that invite children to perform math problems, such as asking them to figure out how many batteries are needed to turn on the lights on a Christmas tree. Children then need to solve the math problem to fulfill the story. They found providing a purpose or a story for children to relate to (e.g. adding coins to a bank to buy a toy car), may better engage them to solve the math problem. \cite{kang2020armath}. In a different domain, AR and storytelling is used to support students’ learning of a socio-scientific issue on nuclear energy use and radiation pollution. Students were asked to imagine their campus being 12km away from the Nuclear Power Plant, and they were on the first day after the hydrogen gas exploration. These ninth-graders then use AR-embedded Android tablet computers to collect data of simulated radiation values on their campus. They found AR can possibly affect learners’ affective attitudes toward real-world issues. Storytelling, game-based learning and problem-based learning might go hand in hand, many AR games included problem-solving features in the design \cite{wu2013current,squire2007augmented}.

\textbf{Learning by doing.} A great number of XR systems seek to facilitate the process of learning by making or learning by doing, by bridging and connecting virtual and physical worlds, e.g. \cite{kaufmann2000construct3d,martin2017virtual,kerawalla2006making,muller2017learning}. Specifically, these systems frequently provide virtual scaffolding, hints and richer information as learners engage in physical tasks . As one example, AR PhonoBlocks \cite{fan2018tangible} is an app that teaches rural county students the alphabetic principle of English. The key features are overlaid dynamic colour cues on 3D physical letters. Specifically, when learners place letters on a surface, the system can dynamically recognize which letter (pair) has been placed, and color-code the letter (pair) regularly pronounced together. Additionally, a 3D simulation of relevant pictures were used to augment the word to be learned to render the learning process more visually engaging (e.g. when children correctly piece together the word “crab”, a 3D simulation of crab will show up). Similarly, in CS education, MR systems were used to turn the placement of physical cards/ tiles into runnable commands, to teach children coding concepts \cite{kim2019mixed, sakamoto2019code}. Similar idea to augment the learning by doing process with XR virtual content is also used in math education in teaching non-symbolic math
\cite{becsevli2019mar}.  

\textbf{Role-play and Participatory Simulation.} “Participatory simulation” is defined as allowing “different players to function as interacting components of a dynamic system” and consequently interactions among students affect the outcomes of the system \cite{klopfer2008augmented,wu2013current}. In educational XR systems, they frequently involve learners to play the role of an object/character related to the target knowledge, and is frequently adopted in educational XR activities, e.g.\cite{keifert2017agency,lindgren2011supporting, allison2000virtual,dunleavy2009affordances,squire2007augmented,klopfer2008augmented} These activities could go hand-in-hand with collaborative learning (e.g. Jigsaw) \cite{dunleavy2009affordances,klopfer2008augmented,squire2007mad,squire2007augmented}. For example, to teach children the state of matter, a collaborative MR environment named STEP allows students to play the role of water particles. The way children move changes the state of the water (e.g. liquid, solid), thus they need to work collaboratively and coordinate movement to form the target water state \cite{keifert2017agency}. In another VR system, learners play the role of juvenile gorilla and interact with other virtual gorillas, to learn about gorilla’s acceptable behaviors, habitat and norms etc. \cite{allison2000virtual}. Research found role-play and “body-based metaphors” can lead children to better grasp the “deep structure” of the learning domain. In the cases when the actual experiences/ scenarios are not physically accessible (i.e. too dangerous, costly or unethical), such role-playing serves as a viable alternative.

\subsubsection{Informal learning}
Informal learning is normally defined as unplanned, without set-goals and self-directed, which happen more naturally, sometimes unintended, and usually occurs outside of a conventional learning setting. As an example, \cite{birchfield2008mixed} used MR to build interactive artwork that people can interact with, enjoy and explore in the art museum context. The self-directed learning in museums is typically considered as informal learning, as it is more exploratory and normally without clear goal or learning outcome measures. Informal learning also usually concern knowledge that is “nice to know”, but not necessarily connected with learners certification or degree. One example is FingAR Puppet that uses AR-enhanced finger puppet to promote 4-6 years old children’s ability of reasoning about emotional states, communication and divergent thinking \cite{bai15blackwell}.

\subsubsection{Discussion}
Researchers as \cite{radianti2020systematic} argued the true potential of VR lies not in better teaching of declarative knowledge, but in offering opportunities to ‘‘learn by doing’’ which is often very difficult to implement in traditional lectures. However, we realize for practitioners designing XR systems, it may be that learning settings are already somewhat predefined. Still, it may be worth noting that the difference between formal and informal learning are not always clear, nor are they mutually exclusive. If practitioners are trying to decide whether to choose formal or informal learning activities, besides considering technologies , several instruction-driven and learner-centered guiding questions could be helpful to consider, including: 1) Does the learning outcome need to be measurable? 2) How self-motivated are your target learners? 3) How capable are your learners of self-directed learning (which informal learning may ask for)? 4) What degree of gamification is appropriate for target learners (e.g. depending on their age and motivation) and content knowledge?

\begin{figure}[htbp]
\centering
\includegraphics[width=0.85\columnwidth]{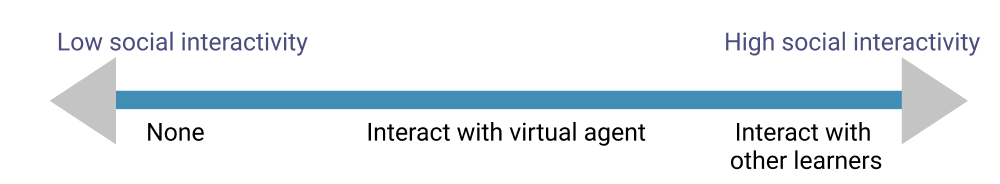}
\caption{The spectrum of D3: Social Interactivity}
\label{fig:d3}
\end{figure}

\subsection{D3: Social Interactivity -- How much support for social interaction are there in the system?}

The third dimension concerns the system's support for social interactivity, broadly defined as the amount of communication learners have while engaging in the systems. Educational XR systems can have various degrees of social interactivity built in with different subjects (e.g. agent or human learners). 

\subsubsection{None}
Some XR applications have little or no support for social interaction for simplicity. They may serve more as the purpose of toolkit rather than an interactive learning environment. As one example, Construct3D is a VR app for mathematics and geometry education, aiming to be a simple and intuitive 3D construction tool in an immersive virtual environment \cite{kaufmann2000construct3d}. XR apps used in formal learning settings (e.g. classroom) that mainly serve the purpose of supplementing formal instructions with some 3D visualization or tangible physical objects, may also have little or no support for social interaction, e.g. \cite{chen2015employing,turan2018impact}. 

\subsubsection{Interact with virtual agent}
XR systems that have a virtual agent that learners can interact with or talk to, can increase the social interactivity that system affords. A virtual agent can guide learners to progress through the learning activities (i.e. relating to Signaling Principle), provide context, background knowledge or storyline (i.e. storytelling), to engage learners in the learning tasks. Such an agent could be human-like or non-human-like (e.g. animal, monster) and normally 1) have functions of verbal communication, 2) are friendly and non-intimidating (especially if interact with young children); 3) have a name that target learners could call and relate to; 4) serve to build connection between the learners and the content, so they may initiate conversation with reference to the learning materials, and drawing users’ attention as proper. For example, ARMath has Victor, a friendly virtual “monster” agent to illustrate what daily situations learners are target with to solve mathematically \cite{kang2020armath}. In a system to teach non-symbolic math, children are asked to help Dima (a virtual dinosaur agent), on a ‘magical stone quest’. To that end, the child places objects which make the stones “gain magical powers” and are asked math questions. Upon the correct answer, Dima will give congratulatory audio feedback and will ask children to “reconsider your answer” if incorrect \cite{salman2019exploring}.

\subsubsection{Interact with other learners}
XR (especially AR) can support social learning and collaboration, through enabling learners to interact with the virtual content while engaging in communication in the real world \cite{dunleavy2009affordances,li2020exploring}. For example, a textbook-based AR social learning game for elementary school students to practice math together, in two different modes: 1) competitive mode and 2) collaborative mode, and found learners have higher concentration when they learn with their peers in the competitive mode. \cite{dunleavy2009affordances} adopted a Jigsaw approach that requires students to complete tasks together. Jigsaw approach is a known collaborative learning approach where each learner acquire unique information, and the success of the tasks require each member’s contribution, holding everyone accountable to solve the problem together.

\subsubsection{Discussion}
Research evidence of positive effect on learning generally advocates for social interaction to be built in educational XR systems. The XR systems that give prompt feedback and promote social interaction can promote better students’ engagement by using immersive experiences, reducing distractions \cite{martin2017virtual}. In education, immediate feedback is undoubtedly a golden rule that improves learning. Learners benefit from receiving immediate feedback (e.g. on correctness of their answers or quality of their performance), as compared to doing tasks without feedback \cite{askew2004feedback,dihoff2004provision,epstein2002immediate}. While we do not argue for every XR system to have high degree of social interact prompt feedback to support better learning, which sometimes can be achieved through social interaction with virtual agents or peer learners. In addition, research in e-learning design principles have found people generally absorb information more effectively when they feel there is ‘human’ element included, and when content is personalized, conversational and informal. For XR systems that incorporate virtual agents, it can be beneficial if the agent communicates in a human-like voice as opposed to a robotic voice, use conversational tone. At the same time of building in social interactivity, XR practitioners need to balance the risk of over-scripting or over-explaining \cite{daniel2016moving, dillenbourg2002over}, potential distraction agents or other learners could bring that might interfere with the learning tasks, and avoid extraneous information which may result in learners being cognitive overloaded \cite{wu2013current,chang2006learning}. 

\begin{figure}[htbp]
\centering
\includegraphics[width=0.85\columnwidth]{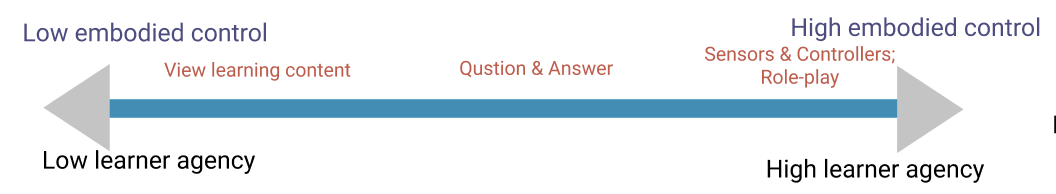}
\caption{The spectrum of D4: Learner Agency}
\label{fig:d4}
\end{figure}

\subsection{D4: Agency -  What level of learner agency do system support?}
The fourth dimension of the XR-Ed framework asks what level of agency learners have. Agency is when learners have the “power to act”, and when learning involves the activity and the initiative of the learners. Higher agency means instead of merely information transmitting to the learners (passive learners’ role), information can also be transmitted from the learners to affect teachers, curriculum and systems (active learners’ role) \cite{hempel2015learner,mercer2011understanding}. In XR settings, learner agency can sometimes go hand in hand with the degree of /textit{embodied control} that systems support (i.e. the degree users can /textit{bodily and physically} interact with the system). 

\subsubsection{Low learner agency}
In these XR applications that promote low learner agency, learners act as passive recipients of knowledge, and participate in the learning activity in a way of reading, viewing, or observing, instead of actively manipulating, learning by doing or playing a part of the XR environment. Examples include in medical education learners view the VR brain images to learn brain anatomy \cite{levinson2007virtual}, going through paintings in a museum that displays them in an immersive virtual reality environment to learn art history  \cite{cecotti2020virtual, cortiz2017web}.

\subsubsection{High learner agency}
Higher agency could give a more immersive experience, which can potentially better engage learners in the learning process. Learning activities that seek more learner input, for example, role-playing, participatory simulation and learning by doing may promote higher learner agency. Examples include the virtual reality (VR) field trips which invite learners to role-play as corals to learn about climate change \cite{markowitz2018immersive}, the MR learning environment in which children role-play as water particles to learn collaboratively the state of the matter \cite{keifert2017agency}, and to improve users’ safety awareness and ability to identify accidents in industrial settings, by allowing them to move around, pick up and manipulate tools, and explore a full range of physical interaction \cite{carruth2017virtual}. 

\subsubsection{Discussion}
While learners could still actively construct knowledge in low-agency, non-interactive objects and videos in XR should be used judiciously, and practitioners are encouraged to make the 3D VR content manipulatable and interactive to promote higher learner agency, as learners may generally expect to learn with higher agency (frequently in form of embodied control), when using immersive technologies. This was vividly shown by the 2 by 2 randomized trial conducted by \cite{johnson2020embodied}, that found in VR settings, the group with higher embodied control (and intuitively, higher learner agency) learned more content than the passively observing group with lower embodied control/ learner agency. Yet the different degree of embodiment did not result in a significant difference in learning content in 2D PC, potentially due to different learners expectations - they may expect more agency in a VR environment, but are more accustomed to viewing content passively on PC.

In XR technologies, while giving learners more agency may encourage students to be active learners, permitting autonomous exploration, learning by doing which support a constructivist approach of learning \cite{wu2013current,martin2017virtual,garzon2019systematic}, the notion of agency isn’t simply about handing control over to learners, but involved teachers and technologies designer to create the context that actively involve learners in the moment.

\begin{figure}[htbp]
\centering
\includegraphics[width=0.85\columnwidth]{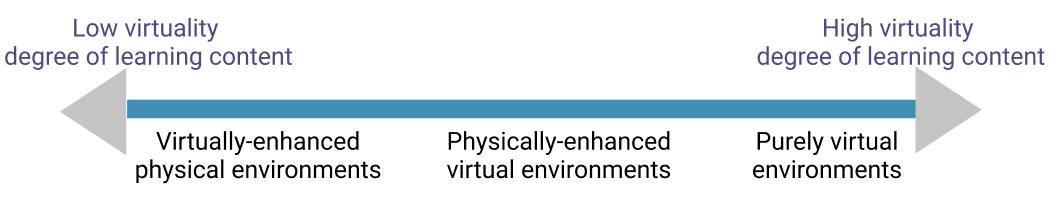}
\caption{The spectrum of D5: Virtuality Degree}
\label{fig:d5}
\end{figure}

\subsection{D5: Virtuality Degree: What degree of reality and virtuality exist in the learning environment?}
With the same learning content, designers can still decide how much virtuality the technology is going to provide in the learning experience. For example, to teach programming concepts along with computational/critical thinking and problem-solving skills, technology designers’ decisions on the virtuality degree can lead to different educational systems and learning experiences described as follows.

\subsubsection{Purely virtual environments} Purely virtual learning environments provided by VR have the unique affordances of immersive technology. \cite{chandramouli2014fun} designed a Virtual Reality (VR) environment for ‘fun-based’ interactive programming instruction in engineering education courses. The system utilizes 3D visualization and immersive interaction to enable students’ more complex logic reasoning which tends to be one of the challenges in learning programming. In terms of the design recommendations of virtual learning material, research showed that the contextually embedded instruction texts can lead to learners’ lower cognitive load and higher self-efficacy, audio is less cognitively demanding as textual information, but text makes information more available for learners \cite{baceviciute2020investigating}. 

\subsubsection{Physically-enhanced virtual environments}
In a physically-enhanced virtual learning environment, the main learning contents are still supported by  virtual objects and virtual interaction, while physical objects enable tangible interaction \cite{kim2019mixed} and provide affordance for contextual learning and scaffolding \cite{draxler2020augmented, kang2020armath}. By bridging the physical world and virtual world by leveraging AR/MR technology, researchers can enable students to physically interact with programming and computational concepts \cite{kim2019mixed}. In ARMath \cite{kang2020armath}, a mobile AR system, researchers utilize the virtual world on the screen to visualize abstract concepts, engage students in virtual math situations with vivid avatars, dialogs for storytelling, and provide different types of scaffolding in time. Meanwhile, physical objects allow tangible interaction with everyday objects, enact real-life practices.. Compared to a purely virtual environment, involving physical objects potentially allow educators to tailor the learning environments to instructional objects in a more accessible and flexible way, such as by tailoring the physical game board without modifying the AR system development \cite{kim2019mixed}.

\subsubsection{Virtually-enhanced physical environments
}
The nuances between virtually-enhanced physical environments and physically-enhanced virtual world are the amount of virtuality used and whether or not the main learning contents are presented physically or virtually. In virtually-enhanced physical learning environments, the physical objects (major learning contents) can be enhanced by the virtually-presented visual cues, scaffoldings, toolbox \cite{fan2020english, jiang2020augmented, liu2020ar, tang2020augmented}, which aim to reduce information overload while students are learning complex knowledge components by providing immediate feedback, in-time learning supports, handy toolkits. Such designs especially can be effective to support lower-level students \cite{tang2020augmented}, children from rural low socio-economic status (SES) schools \cite{fan2020english} or students with lower self-efficacy \cite{liu2020ar}. Virtual augmentation can also provide data visualization to reveal fine details while students are interacting with invisible science phenomena.

\begin{figure}[htbp]
\centering
\includegraphics[width=0.85\columnwidth]{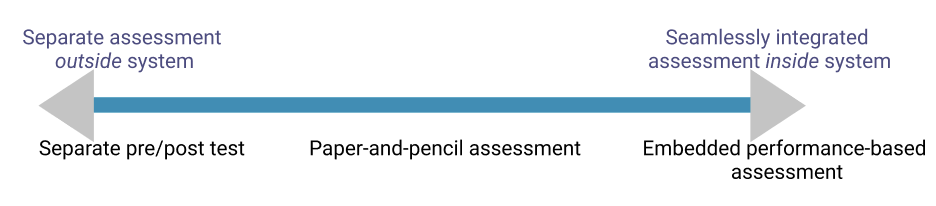}
\caption{The spectrum of D6: Assessment}
\label{fig:d6}
\end{figure}

\subsection{D6: Evaluation of Learning - How seamless are the assessments built in the system?}
The sixth dimensions XR-Ed framework looks at how seamless are the assessment of learning embedded in the XR systems. Assessment is the process of “observing a sample of a student's behavior and drawing inferences about the student's knowledge and abilities” \cite{hurst2015forms}. XR offers potential for various assessment designs, which, depending on how integrated these assessments are in the XR systems, roughly include separate pre/post test, paper-and-pencil assessment, and embedded performance-based assessment. 

\subsubsection{Separate pre/post test outside of the system:}
Many educational XR systems have no learner assessment built in, (e.g. \cite{kang2020armath,turan2018impact,draxler2020augmented}. They rely, rather, on external pre or post tests that assess the learners. For example, in an AR application for college students’ geography education, students were assessed by given a 30-item achievement test as pretest and posttest as assessment \cite{turan2018impact}. Similarly in \cite{draxler2020augmented}'s AR application to learn case grammar, learners were assessed using questionnaire on their prior knowledge of vocabulary and case grammar (as pre-test), and performed vocabulary recall test and transfer performance test (as post-test) to assess their knowledge of case grammar . One design tradeoff is this might be logistically more complicated for learners, but could reduce the development effort and bypass the technical complexity of implementing assessment in XR environment.

\subsubsection{Paper-and-pencil assessment embedded in the system:}
Paper-pencil assessments refer to the traditional way where students provide written responses to items either on paper or on electric forms. Typically, paper-pencil assessments include questions to answer, topics to address through paragraph responses, problems to solve, etc. By this, we refer to traditional methods that ask learners for written responses to demonstrate their knowledge built in the XR systems. This is less commonly observed than the other categories in our review, but one example is in the context of learning art history in the museum, children were quizzed on the art knowledge after going through the VR-enhanced paintings \cite{cecotti2020virtual}. The “learning activities” and “assessment activities” are separate, rather than integrated.

\subsubsection{Performance-based assessment embedded in the system:}
Performance-based assessments refer to students demonstrating their knowledge and skills in a non-written fashion \cite{linn1991complex,darling1994performance}. They focused on demonstration versus written responses. For example, giving oral presentations, completing physical assessments in physical education (PE) classes, performing experiments in a lab, or dissecting activities in anatomy classes fall under this category \cite{hurst2015forms}. Here by embedded performance-based assessment we mean “assessments that ask students to demonstrate their ability through interaction with the system”. Instead of pausing from learning activity to take a test, learning activities can collect user interaction data to assess learners’ learning. One example is, in a VR game that teaches children about natural selection, users play the VR game by manipulating a virtual net to catch the correct species of butterflies \cite{johnson2020embodied}. Whether the learners caught the required number of butterflies was a proxy for learners’ mastery of the knowledge. Similarly, in a mixed reality to teach children non-symbolic math, learners place objects “gain magical powers” and solve math questions. Their answer will be “evaluated by” virtual dinosaur (Dima), who will also give immediate corrective feedback \cite{salman2019exploring} When “learning activities” and the “assessment activities” are not separated but deeply integrated, learners do not need to switch back and forth, and the learning experience can happen in a smoothie way, as learners are “unconsciously” assessed through the relatively more enjoyable learning activities.

\subsubsection{Discussion}
Assessment is an integral part of instruction, as it determines whether or not the goals of education are being met. Assessment affects decisions about grades, placement, advancement, instructional needs, curriculum, and, in some cases, funding. Assessment inspires education practitioners to ask hard questions: "Are we teaching what we think we are teaching?" "Are students learning what they are supposed to be learning?" "Is there a way to teach the subject better?" Assessment not only helps students see their performance and determine their own understanding, but also helps teachers to assess and reflect if their instruction meets the desired goal and learning objectives \cite{kanwar2012assessment,wiggins2005understanding}. 

Somewhat unintuitively, as prior work noted, a lot of educational XR applications do not measure learning outcome, but focus their evaluation more on system usability testing \cite{radianti2020systematic}. This may be improved as the domain matures, but researchers note future educational XR applications should be “more thoroughly evaluated by employing quantitative and qualitative research methods to assess the students’ increase of knowledge and skills as well as the students’ learning experience” \cite{radianti2020systematic, ibanez2018augmented}.

\section{Discussion}
\subsection{Summary of Review Result and Current Barriers}

Our review included work on Virtual Reality (VR) \cite{abadia2018salient,baceviciute2020investigating,cecotti2020virtual,gelsomini2016affordable,hirabayashi2019development,johnson2020embodied,kaufmann2000construct3d,koirala2019exploration,saunier2016designing,shen2017behavioral}, Augmented Reality (AR) \cite{ashtari2020creating,bai15blackwell,chen2015employing,cheung2020techniques,draxler2020augmented,fan2020english,jiang2020augmented,kang2020armath,li2020exploring,liu2020ar,radu2020relationships,radu2019can,turan2018impact,villanueva2020meta} and Mixed Reality (MR) \cite{becsevli2019mar,birchfield2008mixed,fan2020english,hirabayashi2019development,keifert2017agency,khan2018mathland,kim2019mixed,lindgren2011supporting,liu2007mixed,sakamoto2019code,salman2019exploring,vazquez2017serendipitous,yannier2015learning} for education, which we make explicit effort to include a balanced number of each category. Work covered in this review a variety of XR systems from the learning standpoint. For example, for the education setting, the reviewed work span from informal learning \cite{baceviciute2020investigating,bai15blackwell,cheng2017teaching,draxler2020augmented,kang2020armath,keifert2017agency,li2020exploring,yannier2015learning}to formal learning \cite{chen2015employing,fan2020english,jiang2020augmented,liu2007mixed,liu2020ar,turan2018impact}, the target learners age range from K-12 children \cite{chen2015employing,cheung2020techniques,fan2020english,jiang2020augmented,li2020exploring,liu2020ar,salman2019exploring} to college students in higher ed settings or other adults \cite{baceviciute2020investigating,cecotti2020virtual,draxler2020augmented,johnson2020embodied,kaufmann2000construct3d,khan2018mathland,turan2018impact,villanueva2020meta}; the domain subjects that educational XR systems target in our review also span from math \cite{kang2020armath,kaufmann2000construct3d,khan2018mathland,li2020exploring}, science (e.g. physics \cite{liu2020ar,radu2020relationships,radu2019can,yannier2015learning}, animal science \cite{allison2000virtual}), programming and coding \cite{kim2019mixed,sakamoto2019code}, to language learning \cite{cheng2017teaching,cheung2020techniques,draxler2020augmented,fan2020english,hirabayashi2019development,vazquez2017serendipitous} and art \cite{birchfield2008mixed,cecotti2020virtual}. The general review method of this paper was inspired by a review on teacher augmentation systems \cite{an2020ta}.

Through our review, though immersive technologies hold great potential for education, research found many technological and usability barriers (such as dizziness and cumbersome devices) (Wu et al. 2013; Parmaxi and Demetriou 2020; Radianti et al. 2020; Martín-Gutiérrez et al. 2017). Recently, Radianti et al identified  pedagogical issues, such as resistance from schools and teachers, a lack of clear instruction design guidelines, inflexibility of AR systems for customizing instructional content; as well as learning issue, such as XR may cause students to be cognitively overloaded, overwhelmed or confused. As a potential solution, we additionally outlined an instructional design guideline for helping guide practitioners with instructional design in XR technologies. 

\subsection{Step-by-step Approach for Designing XR Instructional Activities}
Based on the XR-Ed Design Framework, we incorporate known theoretical approaches, \textit{Understanding by Design and Backward Design}, enriched with learning sciences or participatory design principles, to offer a step-by-step instructional design approach to guide practitioners’ design process of educational XR instruction.

\subsubsection{Step 1: Identify desired results (Identify Learning Goal)}
For this stage, designers should clearly identify what they aim to enable students to to \textit{know}, to \textit{understand}, and to \textit{do} \cite{wiggins2005understanding}. When selecting learning materials, practitioners should consider materials that are anchored in real world problems that matter to the learners (Anchored learning) \cite{graesser2009inaugural}. 

\subsubsection{Step 2: Determine acceptable evidence (Assessment Design)}
Perhaps unintuitively to most people, \textit{Backward Design} suggests that designers should start to plan out what \textit{assessment} before planning out instruction. As this approach advocates using desired learning result (end goal) to guide the design of instruction, (thus the name ‘backward’), this necessitates placing assessment design prior to instruction design. In this stage, designers should consider if and how the assessment can be embedded into the XR learning system (D6 in XR-Ed Design Framework). In deciding and planning out form of assessment, we note some educational principles that could be helpful for practitioners to keep in mind 1) It is better for students’ learning if they expect an assessment (i.e. avoid “pop quiz”); 2) a spaced schedule of studying and testing (as opposed to all at once) is usually better for learning (Spacing Effect); 3) when deciding the difficulty of tests, they should take into students’ skill level and prior knowledge (Goldilocks Principle) \cite{graesser2009inaugural}. 

\subsubsection{Step 3: Plan learning experience and instruction (Instruction Activities Design)}
Having identified results and assessment methods, it is time to design appropriate instructional activities. In this step, practitioners could go to D2,D3,D4,D5 in the XR-Ed Design Framework to examine the design space, look for design inspirations from existing XR systems, and consider the some design recommendation or tradeoff we noted. 

Additionally, some general evidence-based principles that practitioners could consider at this stage include: 1) consider using stories or example cases, as they tend to be more effective than didactic facts and abstract principles (Stories and Example Cases)\cite{graesser2009inaugural}; 2) eliminate extraneous information on XR media to keep learners attention focused on the most central learning content (Coherence Principle) \cite{clark2003Elearing}; 3) Actively involve academic staff to co-designing instructional activities with them (e.g. teachers, instructional designers, even students). Co-designing educational technologies with stakeholders who are to use it in actual daily practice (e.g. classroom teaching) is the most direct way of gauging what they care about and what kind of systems would be most beneficial for them in their daily activities \cite{roschelle2006co,penuel2007designing, holstein2019co}. Having educational technology more aligned with stakeholders’ (i.e. teachers) preference, could potentially reduce resistance from school teachers to adopt XR technologies and maximize learning benefit \cite{holstein2019co, radianti2020systematic}. 

\begin{table}[]
\begin{tabular}{|l|l|l|}
\hline
\textbf{Step in XR-Ins Guideline} & \textbf{\begin{tabular}[c]{@{}l@{}}Dimension to consider in \\ XR-Ed Framework\end{tabular}}     & \textbf{\begin{tabular}[c]{@{}l@{}}Example Relevant \\ Learning Sciences Theories\end{tabular}}  \\ \hline
Identify Learning Goal            & D1 - Accessibility                                                                               & Anchored learning                                                                                \\ \hline
Assessment Design                 & D6 - Assessment                                                                                  & \begin{tabular}[c]{@{}l@{}}Exam Expectation\\ Spacing Effect\\ Goldilocks Principle\end{tabular} \\ \hline
Instruction Activities Design     & \begin{tabular}[c]{@{}l@{}}D2 - Scenario\\ D3 - Social Interactivity\\ D4 - Agency\end{tabular} & \begin{tabular}[c]{@{}l@{}}Stories and Example Cases \\ Coherence Principle\end{tabular}        \\ \hline
\end{tabular}
\label{tab:step}
\caption{Mapping between different step in XR-Ins Guideline to dimensions in XR-Ed Framework, and example corresponding educational principles}
\end{table}

\section{Conclusions and future work}
We have introduced XR-Ed Framework, which aims to 1) reveal the underlying, sometimes implicit design choices in educational XR applications, 2) provide design inspirations for XR designers and developers by discussing some design implications and trade-offs, 3) propose a common lens where future XR researchers can analyze educational XR systems. Additionally, we outlined XR-Ins design guidelines, a step-by-step instruction-centered design guidelines, \cite{wiggins2005understanding}, each step connecting with the XR-Ed design Framework, enriched by educational principles. 
Our generalization of the XR-Ed framework is not meant to be exhaustive, nor is each design dimension exclusive and separate from each other. Given the rich design space and numerous XR systems with a wide range of designed features, it is likely zmore design dimensions, and we welcome researchers to comment, refine and jointly contribute to comprehend this design space. 

While we tried to clearly define each dimension, some of the dimensions coordinate can be considered non-definitive, and may themselves be better defined in spectrums rather than in category. For example, there is sometimes ambiguity to classify a given learning scenario into either formal learning or informal learning in a very definite way (e.g. an educational game in K-12 classroom setting), as these categories may not be mutually exclusive. Thus a more effective way of visualizing and representing the spectrums (e.g. a  three-dimensional one) could be future direction.
In sum, XR-Ed design framework and XR-Ins Design Guideline are jointly proposed to facilitate XR practitioners (i.e. designers, researchers and developers alike) to identify implicit design choices, gather design inspirations from existing systems’ design, and ultimately design and implement more effective learning environment in a more structured, learner-centric way.

\bibliographystyle{unsrt}
\bibliography{main}

\newpage
\appendix
\section{Appendix}

\begin{table}[htbp]
\begin{tabular}{|l|l|}
\hline
\textbf{Mechanisms of XR technologies in Education}                          & \textbf{\begin{tabular}[c]{@{}l@{}}Example Work\\  in XR system\end{tabular}}                                                                         \\ \hline
Use 3D visualization to teach in a clearer or more visually appealing way    & \cite{villanueva2020meta,jiang2020augmented}                                                                                         \\ \hline
Connect physical and virtual world to support learners learning by doing     & \cite{liu2020ar,fan2020english, radu2020relationships}                                                                               \\ \hline
Augment physical objects to provide richer digital information               & \cite{draxler2020augmented}                                                                                                          \\ \hline
Role-play or participatory simulation as part of the target knowledge        & \cite{keifert2017agency,lindgren2011supporting, allison2000virtual,dunleavy2009affordances,squire2007augmented,klopfer2008augmented} \\ \hline
Embedding interactive agent in XR to promote Social-Emotional Learning (SEL) & \cite{salman2019exploring,kang2020armath}                                                                                            \\ \hline
Afford learner control and agency through XR controller, sensors             & \cite{carruth2017virtual,keifert2017agency,markowitz2018immersive}                                                                   \\ \hline
\end{tabular}
\caption{Mechanisms of XR technologies in Education}
\label{Tab:mecha}
\end{table}

\end{document}